# Thermal Conductivity Inhibition in Phonon Engineered Core-Shell Cross-Section Modulated Si/Ge Nanowires

Denis L. Nika[1,2,*], Alexandr I. Cocemasov[2], Dmitrii V. Crismari[2] and Alexander A. Balandin[1,3,*]

[1]Nano-Device Laboratory, Department of Electrical Engineering, Bourns College of Engineering, University of California – Riverside, Riverside, California 92521 U.S.A.

[2]E. Pokatilov Laboratory of Physics and Engineering of Nanomaterials, Department of Theoretical Physics, Moldova State University, Chisinau, MD-2009, Republic of Moldova

[3]Materials Science and Engineering Program, Bourns College of Engineering, University of California – Riverside, Riverside, California 92521 U.S.A.

**Abstract**

We have shown theoretically that a *combination* of cross-section modulation and acoustic mismatch in the core-shell Si/Ge nanowires can lead to a drastic reduction of the thermal conductivity. Our calculations, which utilized two different models – five-parameter Born-von Karman and six-parameter valence-force field – for the lattice vibrations, indicate that the room temperature thermal conductivity of Si/Ge cross-section modulated nanowires is almost *three orders* of magnitude lower than that of bulk Si. Thermal flux in the modulated nanowires is suppressed by an *order of magnitude* in comparison with generic Si nanowires. The effect is explained by modification of the phonon spectra in modulated nanowires leading to decrease of the phonon group velocities and localization of certain phonon modes in narrow or wide nanowire segments. The thermal conductivity inhibition is achieved in nanowires *without* additional surface roughness and, thus, potentially reducing degradation of the electron transport. Our results suggest that the acoustically mismatched cross-section modulated nanowires are promising candidates for thermoelectric applications.

[*] Corresponding authors: (DLN) dnica@ee.ucr.edu and (AAB) balandin@ee.ucr.edu

Spatial confinement of acoustic phonons in semiconductor thin films and nanowires (NWs) can change their properties in comparison with the corresponding bulk materials [1-4]. Phonon confinement and subband quantization in nanostructures lead to modification of the phonon density of states (DOS) [5-10], electron – phonon scattering rates [10-16], optical response of the materials [17], and phonon relaxation via scattering on defects and in anharmonic Umklapp processes [5, 7-8, 18-20]. It was predicted theoretically within the continuum approximation for phonons that the electron mobility can be increased in phonon engineered core-shell NWs or planar heterostructures via utilization of the barrier shell materials, which are acoustically harder than the core material [21-22]. Experimentally observed reduction of the thermal conductivity, κ, in thin films and nanowires is usually attributed to increased phonon – boundary scattering when the lateral dimensions of nanostructures become comparable to the phonon mean free path (MFP) in a given material. However, another mechanism of κ reduction related to phonon spectrum modification and decrease of the phonon group velocity was also proposed theoretically for the quantum wells [5, 18, 23] and nanowires [7-9, 19-20, 23]. The prediction of the two-orders of magnitude decrease of κ of Si nanowires at room temperature (RT) [20] found experimental confirmation [24] although the exact mechanism remained elusive.

The early results for the phonon spectrum, group velocity and κ in thin films and nanowires obtained within the elastic continuum approximation [5-6, 9, 18-20, 25] or the face-centered-cubic-cell model for lattice dynamics [23, 26] were confirmed later with independent molecular dynamics simulations [27-29]. Both computational approaches gave similar results and indicated that the acoustic impedance $\eta = \rho \times v$ of the barrier shells of nanostructures presents an important tuning parameter for phonon transport, which can be used together with lateral dimensions and shape for phonon engineering of thermal and electron transport in nanostructures ($\rho$ is the mass density and $v$ is the sound velocity of the material). In the core-shell nanostructures with the acoustically mismatched barriers (also referred to as cladding or coating layers) new types of phonon modes (PM) appear: core-like modes, which are concentrated mainly in the nanostructure core, cladding-like modes, which are localized in cladding layers and propagating modes, which extend to both the core and claddings [6, 9, 23, 25-26]. Changing the material and thickness of cladding layers one can decrease the electron – phonon interaction and increase the electron mobility [21-22]. In the same way one can either increase or decrease the phonon group velocity and κ in such nanostructures [6, 9, 23, 25-26].

Recent experimental results added interest to phonon engineering of thermal transport in nanostructures. A two orders-of-magnitude increase in the thermoelectric figure of merit of Si nanowires was attributed to the corresponding reduction of the phonon κ in two independent studies [30-31]. Roughness of nanowire surfaces, which increases the phonon – boundary scattering, was considered to be responsible for the observed κ decrease in one study [31]. Another investigation of κ of Ge and Ge/Si core-shell NWs with the diameters in the range from 15 nm to 20 nm attributed κ reduction specifically to the phonon confinement effects [32]. Dispersion of acoustic phonons in free-standing Si membranes with the thickness of ~8 nm has been measured directly [33]. The authors observed reduction of the phonon group velocities by more than an order of magnitude. Experimentally determined modification of the phonon spectrum was in line with the theoretical predictions for free-standing Si thin films made more than a decade ago [5]. The confined THz PMs with non-zero energy at the center of the Brillouin zone (BZ) were also observed in GaN nanowires using the light scattering measurements [34]. These experimental results confirmed theoretical conclusions about redistribution of the phonon energy spectra in GaN nanowires [9, 35].

Both mechanisms of the thermal conductivity reduction – rough boundary scattering and phonon spectrum modification – can take place depending on the size of the structure, acoustic mismatch and interface quality. The reduction of the thermal conductivity via rough boundary scattering in nanowires can be achieved in nanowires with larger diameter and with less strict technological requirements. However, interface roughness that reduces the thermal conductivity in nanowires also degrades the electron transport. It was recently reported the surface roughness scattering becomes the dominant mechanisms that limits electron mobility in Si nanowires with the rectangular [36] or cylindrical [37] cross-sections.

In this letter we report the phonon engineering approach for κ and thermal flux (TF) inhibition that does not rely on additional roughening of the interfaces. The latter allows one to reduce degradation of the electron mobility [38-39]. Fabrication of cross-section modulated nanowires (MNWs) with layer thickness from several ML up to several nanometers is still a technological challenge. However, recent reports of fabrication of InP, InN/InGaN or metallic MNWs [40-44] suggest that nanostructures considered in this work are feasible.

D.L. Nika, A.I. Cocemasov, D.V. Crismari and A.A. Balandin (2013)

The schematic view of Si/Ge core-shell NW is shown in Figure 1. The Si/Ge NW consist of two periodically repeated Si segments with dimensions $d_x^1 \times d_y^1 \times l_z^1$ and $d_x^2 \times d_y^2 \times l_z^2$ covered by Ge shell with the thickness $d_{Ge}$. To demonstrate the effect of κ suppression in modulated core-shell nanowires we consider here the following modulated NWs and generic NW: Si/Ge MNWs with $14 \times 14 \times N_z - 22 \times 22 \times N_z$ ML Si core and different $d_{Ge}$ (denoted below as MNW # $N_z$ - $d_{Ge}$), Si MNWs with dimensions $14 \times 14 \times N_z - 22 \times 22 \times N_z$ ML (designated hereafter as MNW # $N_z$) and Si NW with the cross-section area $d_x \times d_y = 14 \times 14$ ML, which are equal to the cross-section area of a thin Si segment of MNWs (denoted below as NW #1).

We employed two different models for crystal lattice vibrations: five-parameter Born-von Karman (BVK) and six-parameter valence force field (VFF) models. In both models, phonon energies were calculated from a system of equations of motions written for *all* atoms in the NW or MNW translational period [45-48]:

$$\omega^2 w_i(\vec{r}_n; q_z) = \sum_{j=x,y,z; \vec{r}_m} D_{ij}(\vec{r}_n, \vec{r}_m) w_j(\vec{r}_m; q_z), \quad n=1,\ldots,N, \; i=x,y,z \quad (1)$$

where $D_{ij}$ are the dynamic matrix coefficients describing the interaction between atoms at $\vec{r}_n$ and $\vec{r}_m$, $w_i$ is the $i$th component of the phonon displacement vector, $q_z$ is the phonon wave-number, $\omega$ is the phonon frequency and $N$ is the number of atoms in the NW or MNW translational period. For an atom at $\vec{r}_n$, the summation in Eq. (1) is performed over all nearest and second-nearest atoms at $\vec{r}_m$. The description of BVK and VFF models for 1D and 2D nanostructures were reported by some of us elsewhere [45-48]. In VFF model, we take into account two-, three- and four-particle interactions: stretching, bending, stretching-stretching, stretching-bending and bending-bending. All force constants were obtained from the best fit to experimental dispersion for bulk Si [49] and Ge [50]. The calculations of the phonon thermal conductivities $\kappa_{ph}^{NW}$ and $\kappa_{ph}^{MNW}$ as well as TF per unit temperature gradient $\Theta$ in the NW and the MNWs were performed using the expressions derived in Refs. [45, 51] from the Boltzmann transport equation [7, 45-46, 51-52]:

$$\Theta = \frac{1}{2\pi k_B T^2} \sum_{s=1,...,3N} \int_0^{\omega_{s,\max}} [\hbar\omega]^2 \upsilon_{z,s}(\omega) \tau_{tot,s}(\omega) \exp(\hbar\omega/[k_B T]) \left[\exp(\hbar\omega/[k_B T]) - 1\right]^{-2} d\omega,$$

$$\kappa_{ph}^{NW} = \Theta/(d_x d_y), \quad \kappa_{ph}^{MNW} = \Theta(l_z^1 + l_z^2)/(d_x^1 d_y^1 l_z^1 + d_x^2 d_y^2 l_z^2).$$

(2)

Here $\upsilon_{z,s} = d\omega_s/dq_z$ is the phonon group velocity along the NW or MNW axis, $k_B$ is the Boltzmann's constant, h is the Planck's constant, $T$ is the absolute temperature, $s$ enumerates the phonon branches and $\tau_{tot,s}^{-1}(\omega) = \tau_{U,s}^{-1}(\omega) + \tau_{b,s}^{-1}(\omega)$ is the total phonon relaxation rate for phonon mode $(s,\omega)$. The relaxation time $\tau_{U,s}$ for the Umklapp scattering is calculated using the formula: $\tau_{U,s}^{-1} = B_s \left(\omega_s(q_z)\right)^2 T \exp(-C_s/T)$ [7-8, 46-48, 53] and $\tau_{b,s}$ is the relaxation time for boundary scattering. In the case of NW, we can re-write it as

$$\tau_{b,s}^{-1}(\omega) = \frac{1-p}{1+p} \frac{|\upsilon_{z,s}(\omega)|}{2} \left(\frac{1}{d_x} + \frac{1}{d_y}\right). \tag{3}$$

In the case of MNW it takes the form:

$$\tau_{b,s}^{-1}(\omega) = \sum_{i=1,2} 1/\tau_{b,s}^{(i)}(\omega),$$

$$1/\tau_{b,s}^{(i)}(\omega) = \begin{cases} \frac{1-p}{1+p} \frac{|\upsilon_{z,s}(\omega)|}{2} \left\{ \xi_{core,s}^i \left(\frac{1}{d_x^i} + \frac{1}{d_y^i}\right) + \\ + \xi_{shell,s}^i \left(\frac{1}{d_x^i + d_{Ge}} + \frac{1}{d_y^i + d_{Ge}}\right) \right\}, & \text{if } \xi_{core,s}^i / \xi_{shell,s}^i \geq \delta, \\ \frac{1-p}{1+p} |\upsilon_{z,s}(\omega)| \frac{\xi_{shell,s}^i}{d_{Ge}}, & \xi_{core,s}^i / \xi_{shell,s}^i < \delta, \end{cases} \tag{4}$$

where

$$\xi_{core,s}^i = \int_{-d_x^i/2}^{d_x^i/2} \int_{-d_y^i/2}^{d_y^i/2} \int_{(i-1)l_z^1}^{l_z^1+(i-1)l_z^2} |\vec{w}_s(x,y,z;\omega)|^2 \, dxdydz,$$

$$\xi_{shell,s}^i = \int_{-(d_x^i+d_{Ge})/2}^{(d_x^i+d_{Ge})/2} \int_{-(d_y^i+d_{Ge})/2}^{(d_y^i+d_{Ge})/2} \int_{(i-1)l_z^1}^{l_z^1+(i-1)l_z^2} |\vec{w}_s(x,y,z;\omega)|^2 \, dxdydz - \xi_{core,s}^i. \tag{5}$$

Here $p$ is the specularity parameter, which characterizes the quality of interfaces. The parameter $\delta$ is introduced by us to classify different PMs: core-like, cladding-like and propagating. The quantities $\xi^i_{core,s}$ and $\xi^i_{shell,s}$ show the relative portion of the phonon mode $(s,\omega)$, concentrated in the core or shell of the $i$th MNW segment, correspondingly. Eqs (3-4) represent an extension of formula for the phonon – boundary scattering [54] to the case of a rectangular NW or MNW [45, 51].

In Eq. (4) we take into account that the core-like and propagating PMs in Si/Ge MNWs partially scatter at Si/Ge interfaces and outer boundaries while cladding-like (Ge-like) modes with $\xi^i_{core,s} / \xi^i_{shell,s} < \delta$ scatter only at Si/Ge interfaces. The parameter $\delta$ represents a threshold value for a ratio between the integrated phonon amplitudes concentrated in the core $\xi^i_{core,s}$ and in the shell $\xi^i_{shell,s}$ of $i$th MNW segment. In our analysis we used $\delta = 0.1$, which means that in the cladding-like modes more than 90% of lattice vibrations from the $i$th MNW segment occur in Ge-shell, while Si core-region is depleted of phonons. A similar effect of the phonon depletion was theoretically described by some of us for the acoustically-mismatched planar heterostructures, where part of the PMs is pushed out into the acoustically softer layers [25]. The mode-dependent parameters $B_s$ and $C_s$ in the expressions for the Umklapp scattering in Si/Ge MNWs were averaged for the values of the Umklapp scattering parameters in bulk Si and Ge so that $B_s = (\xi^1_{core,s} + \xi^2_{core,s})B_{Si} + (\xi^1_{shell,s} + \xi^2_{shell,s})B_{Ge}$ and $C_s = (\xi^1_{core,s} + \xi^2_{core,s})C_{Si} + (\xi^1_{shell,s} + \xi^2_{shell,s})C_{Ge}$. In the case of Si NW and Si MNW, we have $\xi^1_{shell,s} + \xi^2_{shell,s} = 0$ and $B_s = B_{Si}$, $C_s = C_{Si}$. The bulk values $B_{Si}$ and $C_{Si}$ were taken from Ref. [45] while $B_{Ge}$ and $C_{Ge}$ were determined by comparing the calculated thermal conductivity of bulk Ge with experimental data [55]: $B_{Si} = 1.88 \times 10^{-19}$ s/K, $C_{Si} = 137.39$ K, $B_{Ge} = 3.53 \times 10^{-19}$ s/K, $C_{Ge} = 57.6$ K.

The phonon spectra in Si NW, Si MNW and Si/Ge MNWs were calculated from Eq. (1) for each value of $q_z$ from the interval $[0, q_{z,max}]$, where $q_{z,max} = \pi/a$ for NW and $q_{z,max} = \pi/L_z$ for MNWs, where $a = 0.543$ nm is the lattice constant of Si and $L_z = l^1_z + l^2_z$ is the period of MNW along the $Z$-axis. In Figure 2, we show dependence of the phonon energies on $q_z$ for the Si NW #1 (panel a) and Si/Ge MNW#8-4 (panel b). The energy dispersion for phonon branches $s=1$-5,7,9,11,..,293 of the Si NW#1 and $s=1$-5,25,50,75,..,4150 of the MNW#8-4 are plotted for comparison. The presented spectra were obtained using BVK model. The total number of phonon branches $N_b$ in Si NWs and Si/Ge MNWs equals to the number of atoms in the nanostructure period multiplied by three degrees of freedom: $N_b = 3 \times N$. The number of phonon branches $N_b$=4152 in the MNW#8-4 is substantially larger than $N_b$=294 in the NW#1. Flattening of the phonon dispersion branches in MNW leads to a decrease in the phonon group velocities in MNW in comparison with NW. As a result, the average phonon group velocities in Si/Ge MNWs are substantially lower than those in both Si NW and corresponding Si MNW. The color of the curves shows contribution of a given PM ($s,q_z$) to thermal transport: red (blue) color designates maximum (minimum) participation. Since κ is a sum of contributions of all PMs, we also calculated partial contribution of each mode. In Si NW, PMs with energy $\omega \leq 10$ meV carry 24% of heat while in the Si/Ge MNW#8-4 these modes carry 50% of heat. Therefore, participation of the higher energy PMs in heat transport is substantially reduced in Si/Ge MNWs. Similar results were obtained using VFF model.

The average phonon group velocity $\langle \upsilon \rangle (\omega) = g(\omega) / \sum_{s(\omega)} (d\omega_s / dq_z)^{-1}$ is shown in Figure (2) for the Si NW#1, Si MNW#8 and Si/Ge MNWs #8-4 and #8-10. Here $g(\omega)$ is the number of the PMs with frequency $\omega$. Note that Ge shell reinforces the decrease of $\langle \upsilon \rangle$ making it especially strong for the low- and middle-energy phonons with $\hbar\omega < 40$ meV. Since these phonons are the main heat carriers in semiconductor nanostructures, the strong decrease of their group velocities significantly influences thermal conduction.

The dependence of κ on temperature, T, for Si NW#1, Si MNW#8, Si/Ge MNWs #8-1, #8-3, #8-7 and #8-10 is presented in Figure 3 (a). The results of the calculations with both BVK (solid curves and dashed curve) and VFF (dotted curve) models are presented. The thermal conductivity of Si MNW#8 is lower by a factor of 4.3 – 8.1 than that in Si NW#1, depending on T. Additional strong decrease of κ is reached in Si/Ge MNWs. Increasing the thickness of Ge shell to $d_{Ge}$ = 7 ML leads to κ decrease by a factor of 2.9 – 4.8 in comparison with that in Si MNW#8, and by a factor of 13 – 38 in comparison with that in Si NW#1. The reduction in κ of Si/Ge MNWs is substantially stronger than that reported for core – shell nanowires without cross-sectional modulation [23, 26-28, 32, 47]. In the generic core-shell nanowires, the κ decrease is due to phonon hybridization, which results in changes in the phonon DOS and group velocity. In our core-shell MNWs, κ reduction is reinforced due to localization of some PMs in wider MNW segments. The localization completely removes such phonons from the heat transport. The difference in κ calculated using BVK and VFF models (see Figure 3 (a)) is explained by the differences in the phonon dispersion. It is small, constituting only ~5-8 %, depending on *T*.

A comparison of κ of nanowires with the same cross-sections is sufficient to make a conclusion about nanowires' abilities to conduct heat. A lower κ means lower heat transfer. In the case of nanowires with various cross-sections, a nanowire with the minimum κ may not necessarily possess the minimum thermal transfer properties due to the difference in the cross-section areas. Therefore, in this case, a comparison of TF is more illustrative. The dependence of the ratio of TF at RT for MNWs and Si NW on $d_{Ge}$ is presented in Figure 3 (b) for $N_z$ = 4, 6, 8, 12. Two points for $N_z$ = 20 and 28 at $d_{Ge}$ = 4 ML are also shown. All curves demonstrate a maximum between $d_{Ge}$ = 3 ML and $d_{Ge}$ = 6 ML. The increase in $N_z$ leads to a shift of the maximum to lower values of $d_{Ge}$.

In order to explain the non-monotonic behavior of the ratio curves in Figure 3 (b) we plot the spectral density of TF $\phi(\omega)$ as a function of the phonon energy for Si NW, Si MNW and Si/Ge MNWs with $d_{Ge}$ = 4 ML and $d_{Ge}$ = 10 ML. The RT data is presented in Figure 3 (c), where $\phi(\omega)$ is determined by the equation $\Theta = \int_0^{\omega_{max}} \phi(\omega) d\omega$. In Si MNWs without Ge cladding, TF is strongly suppressed in comparison with Si NW due to redistribution of phonon energy spectra leading to the reduction of the phonon group velocities and localization of PMs in nanowire segments. The influence of Ge shell on the spectral density of MNWs is two-fold: (i) it reinforces the decrease in the spectral density in Si/Ge MNWs as compared with Si MNWs owing to a stronger decrease of the phonon velocities and stronger PM localizations; (ii) it increases TF due to appearance of additional channels for heat transfer through the Ge shell and attenuation of the phonon boundary scattering of propagating and cladding-like PMs. An interplay of these two opposite effects explains the non-monotonic dependence of the spectral density and TF ratio on $d_{Ge}$ shown in Figures 3 (b) and (c).

The low-energy phonons weakly participate in the three-phonon scattering processes due to the restrictions imposed by the energy and momentum conservation laws [46]. For this reason, these phonons are scattered mostly by the nanostructure boundaries. The relaxation time for the low-energy phonons $\tau(\omega) \sim \tau_B(\omega) \sim \Lambda_b / \langle \upsilon \rangle (\omega)$ increases with the energy owing to the decrease of $\langle \upsilon \rangle$, and it increases with $d_{Ge}$ owing to both the decrease of $\langle \upsilon \rangle$ and increase of $\Lambda_b$. Here $\Lambda_b$ is the phonon MFP in the boundary scattering. Therefore, at small phonon energies the spectral density increases with the increasing energy. At the same time, the number of the Umklapp scattering channels also increases with increasing phonon energy and, as a result, the Umklapp scattering becomes the dominant scattering mechanism for $\hbar\omega > 15 meV$ for all considered nanowires. The difference between TF in MNWs and NWs becomes larger with growing $N_z$ and it reaches the maximum values of 10-11 at $N_z \sim 28$ ML – 32 ML (see Figure 3(b)). For $N_z$>32 ML, TF ratio starts to decrease due to redistribution of the phonon energy spectra and heat conduction through Ge shell.

Figure 4 illustrates the effects of the PM localization in Si core or Ge shell (panel (a)) and in narrow or wide MNWs segments (panel b). In panel (a) we plot the dependence of the PM localization in Si core $\xi_{Si} = (\xi_{core}^1 + \xi_{core}^2) \times 100\%$ as a function of the phonon energy. In Si nanowire $\xi_{Si}$ = 100% (dashed line), while in Si/Ge MNW#8-2 (red triangles) and MNW#8-7 (blue circles) $\xi_{Si} \leq 100\%$. As shown in Figure 4 (a), many low-energy PMs in MNWs with $\hbar\omega \leq 15$ meV are concentrated in Ge shell and possess $\xi_{Si} \leq 20\%$. It means that more than 80% of atomic vibrations in these modes occur in Ge shell. The number of the PMs localized in Ge-shell increases with increasing shell thickness. The energy-dependent localization of the PMs in the narrow segments of MNWs $\xi_{narrow} = (\xi_{core}^1 + \xi_{shell}^1) \times 100\%$ is shown in Figure 4 (b) for Si NW (dashed line), Si/Ge MNW#8-2 (red triangles) and MNW#8-7 (blue circles). Approximately, half of the PMs are concentrated in the wide segments of MNW and have $\xi_{narrow}$<30%, i.e. less than 30% of the atomic vibrations of these modes take place in the narrow segments of MNW. The demonstrated localization of certain PMs in Ge shell or wide segments of MNWs is one of the reasons for the strong inhibition of κ in Si/Ge MNWs.

For ideally smooth interfaces when all phonon scattering events are specular $p$ =1. In our calculations we used $p$=0.85, which corresponds to smooth NW surfaces with small average roughness height $\Delta \sim 1$ML. We estimated $\Delta$ from averaging the mode-dependent specular parameter $\tilde{p}(q,\Delta) = \exp(-2q^2\Delta^2)$ [48, 54, 56-57] over all $q$: $p = \int_0^{q_{z,\max}} \tilde{p}(q)dq / q_{z,\max}$. The small roughness of NWs and MNWs interfaces is beneficial for both maintaining high electron mobility [36-37] and for suppression of the phonon heat conduction in MNWs [45]. Some of us have recently shown theoretically that increase of $p$ leads to faster growth of the thermal flux in Si NW than in Si MNWs without claddings [45]. This effect was attributed to the fact that the higher energy PMs in MNW are excluded from thermal transport while they participate in the heat transfer in NWs. As a result the ratio between TF in Si NW and Si MNWs increases with increasing $p$.

The described approach for inhibition of the phonon thermal transport that does not rely on additional surface roughening and can be achieved with relatively smooth interfaces extends the possibilities of the phonon spectrum engineering. It is distinctively different from the thermal conductivity reduction in Si nanowires with amorphous surfaces [58-59] or Si/Ge quantum dots superlattices [60-61]. One should note also that the difference between the acoustic impedances in Si and Ge is not large: $\eta_{Si}=19.6\times10^3$ kg m$^{-2}$ s$^{-1}$ and $\eta_{Ge}=25.9\times10^3$ kg m$^{-2}$ s$^{-1}$. However, it is sufficient for inducing substantial modification to the phonon spectrum and phonon transport. At RT, the 3-6 ML – thick Ge shell suppresses TF in Si/Ge MNW by an order of magnitude in comparison with Si NW and by a factor of two in comparison with Si MNW without cladding. As a result, $\kappa$ of Si/Ge MNWs is almost three orders of magnitude lower than that of bulk Si. We expect that a larger mismatch between the core and shell impedances would lead to stronger inhibition of the phonon transport.

In conclusion, we have shown that a *combination* of cross-section modulation and acoustic mismatch in the core-shell Si/Ge nanowires results in a drastic reduction of the thermal conductivity and thermal flux. The thermal conductivity inhibition is achieved in the phonon engineered nanowires *without* employing additional surface roughness and, thus, preventing degradation of the electron transport.

*ACKNOWLEDGMENTS*

The work in Balandin Group at UCR was supported, in part, by the Semiconductor Research Corporation (SRC) and the Defense Advanced Research Project Agency (DARPA) through STARnet Center for Function Accelerated nanoMaterial Engineering (FAME). DLN, AIC, DVC acknowledge financial support from the Moldova State projects no. 11.817.05.10F and 12.819.05.18F. AIC acknowledges support under the National Scholarship of the World Federation of Scientists.

**Figure Captions**

**Figure 1:** Schematic of (a) a generic Si nanowire and (b) a Si/Ge cross-section modulated nanowire.

**Figure 2:** Phonon energies as a function of the phonon wave vector $q_z$ in (a) a generic Si nanowire with the lateral cross-section area $14\times14$ ML and (b) a cross-section modulated Si/Ge nanowire with $14\times14\times8-22\times22\times8$ ML Si core and 4 ML – thicker Ge shell. The color in the plot shows the relative participation of the PMs in heat transfer: from maximum (red) to minimum (blue). Note, that cross-section modulation combined with coating effectively removes the higher energy phonon branches from heat conduction. (c) Average phonon group velocity as a function of the phonon energy shown for Si NW with the lateral cross-section area $14\times14$ ML, Si MNW with dimensions $14\times14\times8-22\times22\times8$ ML and Si/Ge core-shell MNWs with $14\times14\times8-22\times22\times8$ ML Si core and different $d_{Ge}$.

**Figure 3:** (a) Phonon thermal conductivity as a function of the absolute temperature. (b) Ratio of thermal fluxes in Si NW and Si/Ge MNWs as a function of $N_z$. (c) Spectral density of the thermal flux in Si NW, Si MNW and Si/Ge core-shell MNWs as a function of the phonon energy. Results in (a, b, c) are presented for Si NW with $14\times14$ ML lateral cross-section, Si MNW with dimensions $14\times14\times8-22\times22\times8$ ML and Si/Ge MNWs with $14\times14\times8-22\times22\times8$ ML Si core and different $d_{Ge}$. Dotted curve in (a) shows phonon thermal conductivity calculated using the VFF model of lattice vibrations. The results demonstrate strong suppression of both thermal conductivity and thermal flux in Si/Ge cross-section modulated nanowires. The thermal conduction inhibition is much stronger than previously reported and is achieved without employing surface roughening.

**Figure 4:** Phonon modes localization as a function of the phonon energy in (a) Si core and (b) narrow segment of MNW. The results are presented for Si MNW#8 (dashed line, panel (a)), Si NW#1 (dashed lines, panel (b)), Si/Ge MNW#8-2 (red triangles) and Si/Ge MNW#8-7 (blue circles).

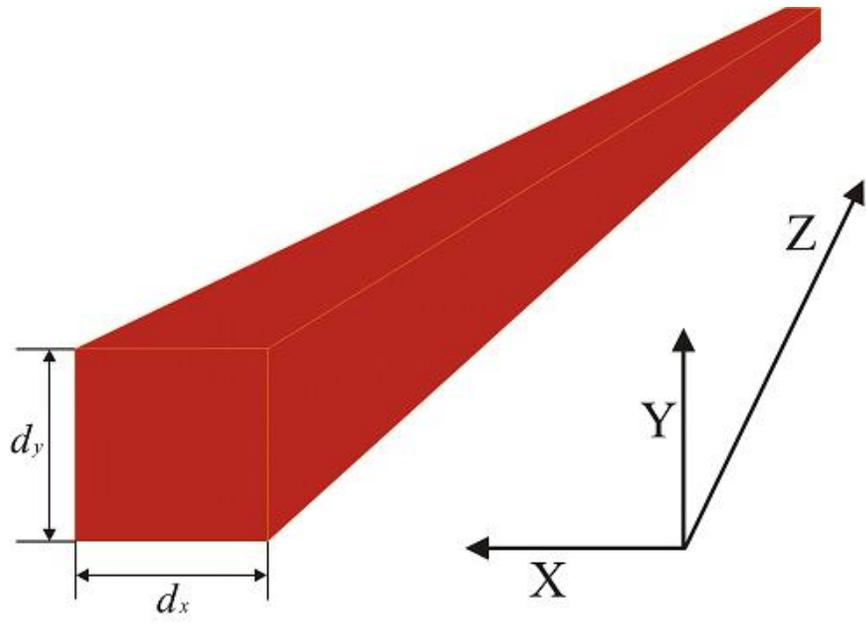

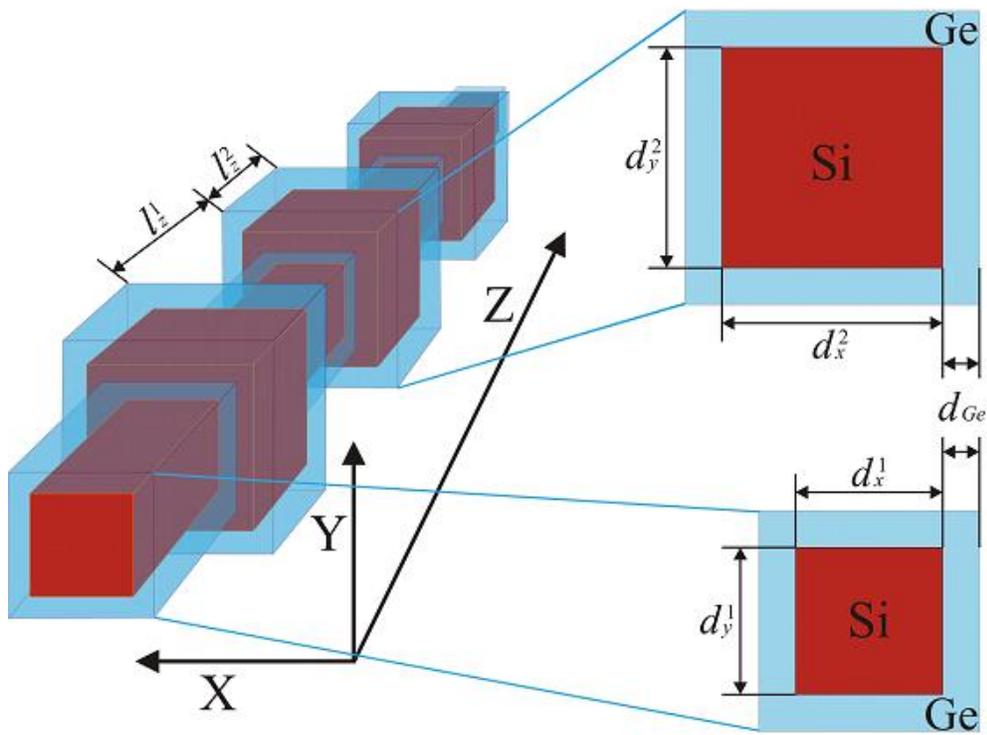

Figure 1

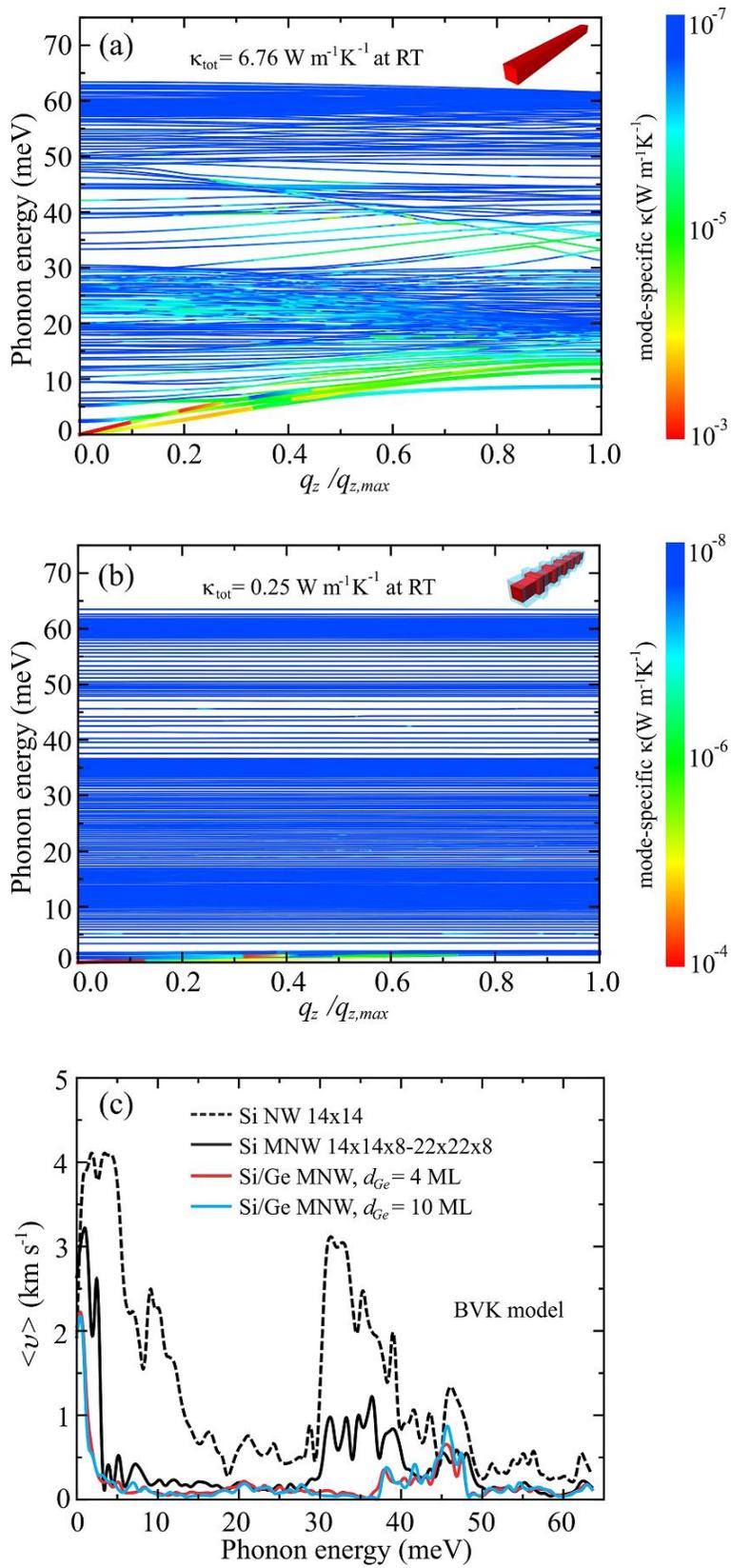

Figure 2

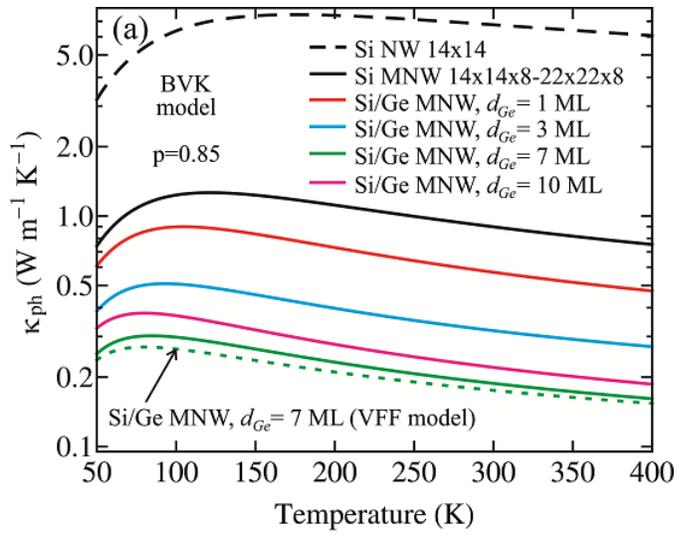
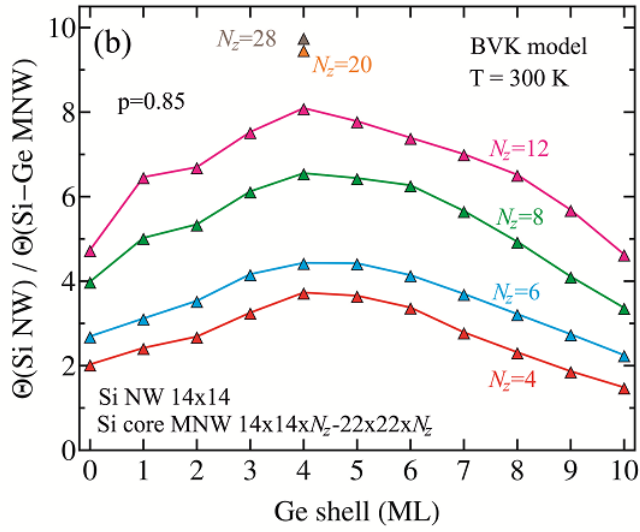
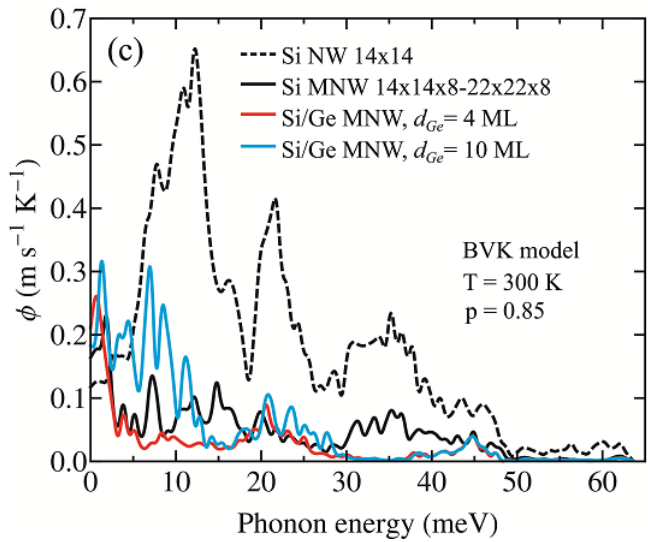

Figure 3

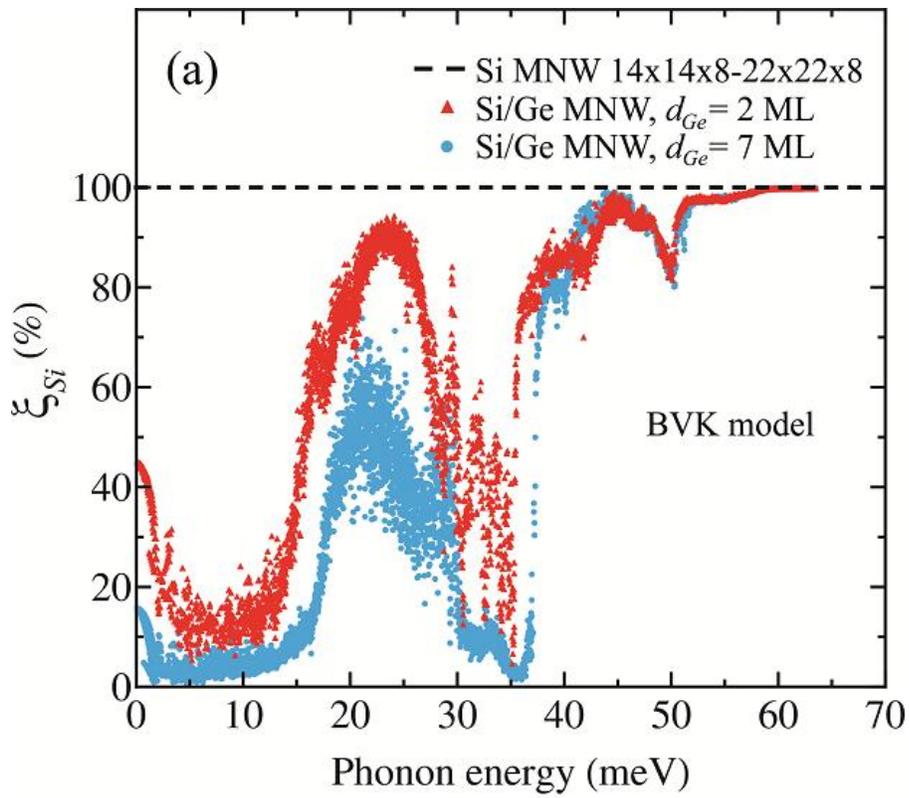

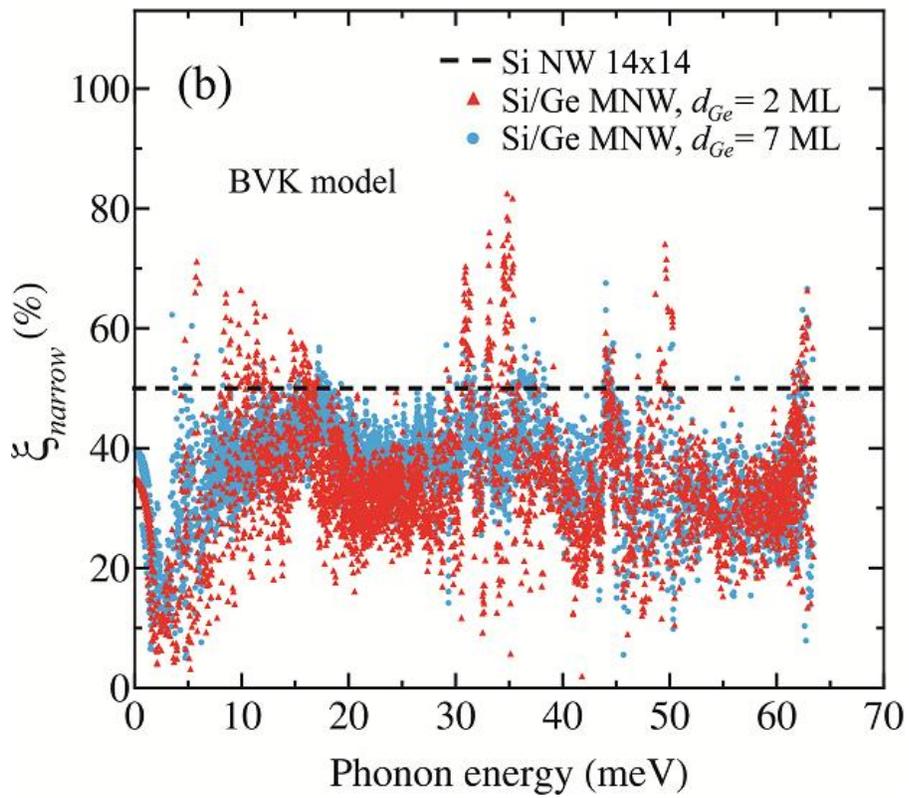

Figure 4